\documentclass[11pt]{article}
\usepackage{lmodern}
\usepackage[T1]{fontenc}
\DeclareMathAlphabet{\mathsf}{OT1}{lmss}{m}{n}
\SetMathAlphabet{\mathsf}{bold}{OT1}{lmss}{bx}{n}
\usepackage{fontspec}
\usepackage[margin=1in]{geometry}
\usepackage{hyperref}
\usepackage{orcidlink} 
\usepackage[blocks]{authblk} 
\usepackage{cite}
\usepackage{amsmath}
\usepackage{amsthm}
\usepackage{dirtree}
\usepackage{fancyvrb,xcolor}

\newcommand{\code}[1]{\texttt{#1}}
\newcommand{\ds}{\displaystyle}
\newcommand{\tsty}[1]{\mathsf{#1}} 
\newcommand{\ra}{\rightarrow}

\newcommand{\infers}{\vdash}
\newcommand{\natpl}{\mathcal{N}_{\textup{\textsf{PL}}}}
\newcommand{\natfol}{\mathcal{N}_{\textup{\textsf{FOL}}}}
\newcommand{\Iimplies}{\textup{I}_{\ra}}
\newcommand{\Inot}{\textup{I}_{\neg}}
\newcommand{\Iand}{\textup{I}_{\wedge}}
\newcommand{\Ior}[1]{\textup{I}_{\vee,#1}}
\newcommand{\Eimplies}{\textup{E}_{\ra}}
\newcommand{\Enot}{\textup{E}_{\neg}}
\newcommand{\Eand}[1]{\textup{E}_{\wedge,#1}}
\newcommand{\Eor}{\textup{E}_{\vee}}
\newcommand{\Iforall}{\textup{I}_\forall}
\newcommand{\Eforall}{\textup{E}_\forall}
\newcommand{\Iexists}{\textup{I}_\exists}
\newcommand{\Eexists}{\textup{E}_\exists}
\newcommand{\Ax}{\textup{Ax}}
\newcommand{\RAA}{\textup{RAA}}
\DefineVerbatimEnvironment{lap}{Verbatim}{
  commandchars=\\\{\},
  formatcom=\setlength{\lineskip}{0pt},
  baselinestretch=1
}
\DefineVerbatimEnvironment{tinylap}{Verbatim}{
  commandchars=\\\{\},
  formatcom=\setlength{\lineskip}{0pt},
  baselinestretch=1,
  fontsize=\fontsize{6}{7}\selectfont
}

\theoremstyle{definition}
\newtheorem{definition}{Definition}[section] 

\title{LAP: Simple Command-line Tools for Teaching Logic, Algorithms,
  and Proof in Computer Science\thanks{To appear at
    \emph{TEAL 2026: Tools for Educational Activities in Logic},
    a FLoC 2026 Workshop, July 25, 2026}}

\author[1]{Stephen F. Siegel\, \orcidlink{0000-0001-9359-3332}}
\affil[1]{%
  Department of Computer \& Information Sciences,
  University of Delaware, Newark DE 19716, USA\\
  Email: \href{mailto:siegel@udel.edu}{\texttt{siegel@udel.edu}}%
}

\author[2]{Yuxin Zhou\, \orcidlink{0000-0003-0456-2011}}
\affil[2]{%
  Department of Computer \& Information Sciences,
  University of Delaware, Newark DE 19716, USA\\
  Email: \href{mailto:sobonlinemsn@live.com}{\texttt{sobonlinemsn@live.com}}%
}

\begin{document}

\maketitle 

\begin{abstract}
  The LAP toolset is a set of command line tools for teaching logic in
  computer science.  It provides implementations of standard
  algorithms for propositional and first order logic, including
  conversions to various normal forms, propositional satisfiability
  algorithms such as DPLL, Tseytin's transformation, and equivalence
  checking.  Significantly, LAP also supports a language for
  expressing a natural deduction derivation for propositional or first
  order logic.  The tools can check the derivation, provide meaningful
  feedback if it is wrong, or display the derivation in a variety of
  formats.  The toolset is written in Java and has no dependencies
  other than a Java Virtual Machine.  The code has been designed to be
  easy to read and to illuminate the data definitions and algorithms.
\end{abstract}

\section{Introduction}

This paper describes certain educational software tools for teaching
logic in computer science.  This toolset is being developed as part of
a curriculum, \emph{Logic, Algorithms, Proof} (LAP), which will
include a text and other material.  It is intended for an advanced
undergraduate or beginning graduate course, and currently covers
propositional and first order logic from a computational point of
view.  The tools supplement the text by providing implementations of
the algorithms described there, including conversion to various normal
forms, Tseytin's
transformation\footnote{\url{https://en.wikipedia.org/wiki/Tseytin_transformation}},
and boolean satisfiability algorithms such as
DPLL\footnote{\url{https://en.wikipedia.org/wiki/DPLL_algorithm}}.
Most importantly, the tools support a language for expressing a
natural deduction derivation, they can check that a derivation is
correct or provide useful feedback when it is not, and can present the
derivation in a variety of formats.

The tools are written in Java and have a simple command-line interface.
They are free, open source, and available at:
\url{https://github.com/verified-software-lab/lap}.

\subsection{Context and Goals}

There are many excellent educational tools for teaching logic and for
checking derivations, including \cite{carnap, logic-machine, fitchFX,
  LPL}. How does LAP differ?  We think the following combination of
features is unique.

First, it emphasizes logic from a computational point of view.  One
way it does this is through the LAP source code itself, which we
intend for students to read.  We have tried to make the code easily
understandable and well-documented.  We strive for simple, natural
implementations that correspond to the pseudo-code in many texts,
rather than highly optimized code.  Using Java also helps, because it
is familiar to many computer science students.

\begin{figure}[t]
  \centering
  \includegraphics[scale=.45]{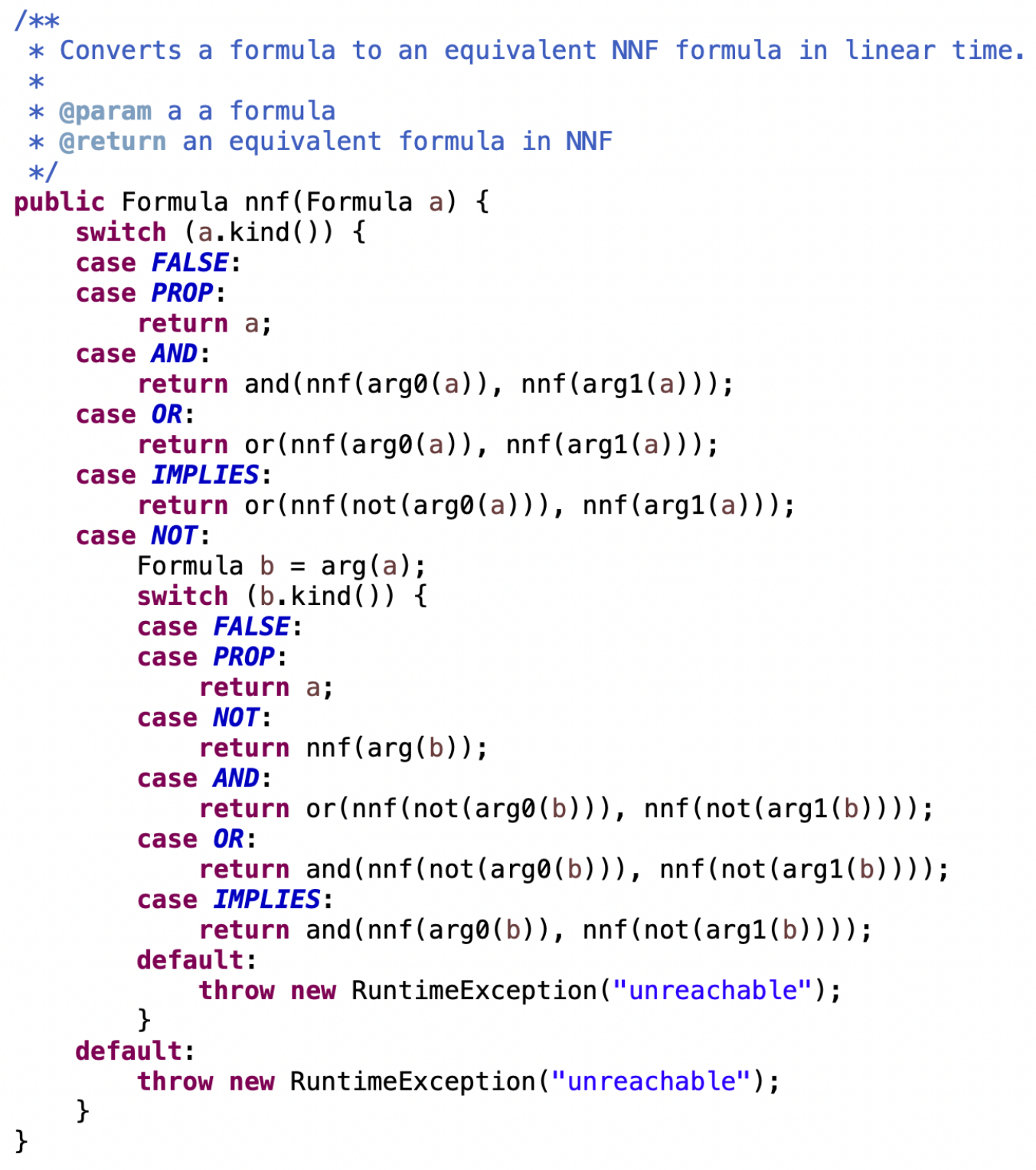}
  \caption{LAP source code excerpt: conversion to negation normal form.
    The LAP code has been carefully developed to be understandable and
    to illuminate the fundamental algorithmic ideas.}
  \label{fig:nnf}
\end{figure}

\emph{Formulas} and \emph{derivations}, two of the fundamental
concepts in logic, are precise mathematical structures that are
naturally represented as inductively defined data types.  In this way
they are similar to lists or trees or similar structures that are
familiar to computer science students.  Some presentations of logic
elide this point, especially for derivations.  In LAP, we emphasize
it.  The \code{Formula} and \code{Derivation} classes look like
typical inductive data type definitions in Java.  Algorithms
manipulating these structures are recursive routines in which the
structure of the code mirrors the structure of the definition, another
familiar concept (cf.\ \cite[\S8.3]{htdp}).  An an example, Figure
\ref{fig:nnf} shows the LAP method for converting a propositional
formula to negation normal form.  Moreover, derivations can be
displayed using a variety of ``views'', e.g., a simple nested tuple
(consisting of a conclusion, rule, and list of subderivations), a
tree, a linear or hierarchical format, or a Fitch
diagram\footnote{\url{https://en.wikipedia.org/wiki/Fitch_notation}}.
Beginning students often mistakenly believe these are different proof
systems, but LAP helps to clarify they are simply different ways of
viewing the same underlying data structure.  Computer science students
are usually familiar with this distinction between data and view,
which they may have seen in an introductory course (e.g., ``world
programs'' in \cite{htdp}) or from studying the Model-View-Controller
design pattern.

As mentioned above, LAP tools are command line tools.  The input and
output (from the terminal or files) consist of plain
text.\footnote{The text may use Unicode characters, color, and other
  highlighting features supported by modern terminals.  Especially in
  the case of derivations, this can render the products visually
  appealing and easier to comprehend.}  This is appropriate for
computer science students.  For example, to construct a derivation, a
user writes a text file using any editor they choose, then invokes
\code{lap check} from the command line.  If LAP finds an error in the
derivation, it provides meaningful feedback.\footnote{The code that
  checks a derivation can also be enlightening for students.  It
  emphasizes another fact that is often glossed over: for a proof
  system to be useful, there must be an effective procedure for
  checking that a candidate derivation is correct.}  The user then
edits the file and tries again, and continues iterating in this way
until the check succeeds.  The process is similar to developing, say,
a C program using an editor and a compiler from the shell.

Our command line tools have several advantages.  There are no dependencies
on remote servers, web browsers, databases, GUIs, IDEs, or other
frameworks (other than a Java Virtual Machine).  A user can easily
incorporate the tools into their own workflow by scripting.  For
example, a teacher could incorporate LAP into a script to grade
student assignments.  In a Unix-style shell, the tools can be easily
combined in interesting ways.  For example, LAP's \code{dpll} command
consumes a propositional formula in conjunctive normal form (CNF) and
determines its satisfiability using the DPLL algorithm.  LAP's
\code{tseytin} command converts an arbitrary propositional formula
into an equisatisfiable formula in CNF.  The two can be composed to
check the satisfiability of an arbitrary propositional formula as
follows:
\begin{verbatim}
> lap tseytin -f '(p<->q) & !(q<->p)' | lap dpll -in 
false
\end{verbatim}
Here, the output from \code{tseytin} is piped to the input of
\code{dpll}.  LAP reports the formula is unsatisfiable.

The command line tools are self-documenting using \code{lap help ...}.
Option \code{-v} instructs LAP to produce not only the final result,
but to show what is going on in an algorithm.  For example, in DPLL,
it displays the state of a depth-first search after each push, pop, or
unit clause reduction.  This helps a student understand the algorithm.
Other options allow formulas or derivations to be read from a file, or
specified as a command line argument, or read from terminal input.

Finally, there are many variations in formalization, proof systems,
notation and terminology across logic texts.  For LAP, we have chosen
to be consistent with the presentation in the book \emph{Rigorous
  Software Development} \cite{RSD}, a leading text on deductive
program verification which also reviews necessary background in
propositional and first order logic. LAP tools are therefore ideal for
students who are using or will use this text.\footnote{A new edition
  of \cite{RSD} is scheduled to be published later this year.}

\section{Basic Usage and Propositional Algorithms}

In this section, we describe the basic propositional algorithms in LAP
and how to access them through the command line interface.

\subsection{The \code{help} command}

Basic command line usage is obtained by typing \code{lap help}:
\begin{verbatim}
> lap help
lap: Logic, Algorithms, Proof tool (language: Propositional Logic)
Usage: lap <command> ...
Commands: 
  help    - print usage information
  nnf     - convert a propositional formula to negation normal form
  cnf     - convert a propositional formula to conjunctive normal form
  dnf     - convert a propositional formula to disjunctive normal form
  tseytin - convert a propositional formula to an equisatisfiable CNF formula
  dpll    - apply the DPLL algorithm to a CNF formula
  sat     - determine if a propositional formula is satisfiable
  valid   - determine if a propositional formula is valid (a tautology)
  equiv   - determine whether two propositional formulas are equivalent
  check   - check a natural deduction derivation

Type "lap help <command>" for detailed help on a specific command.
Type "lap help formulas" for formula syntax.
Type "lap help derivations" for derivation syntax.
Insert "-lang <language>" after "help" to specify language.
Languages: pl (default), fol.
\end{verbatim}

Currently, LAP supports two languages: propositional logic (PL) and
first order logic (FOL).  The default is PL, and the \code{help}
commands show only information relevant to the selected language.
This avoids inundating beginning students with too many unfamiliar
concepts.  A similar pedagogical practice is followed in the \emph{How
  to Design Programs} (HtDP) curriculum \cite{htdp}: students begin
with a very limited programming language (BSL) and the programming
environment reveals only that language.  Students can move to the next
language when they are ready, and then the environment reveals more.

\subsection{Formula syntax}
\label{sec:formulasyntax}

A LAP identifier is as in C or Java: an (upper- or lower-case) letter
or underscore followed by any number of letters, digits, and
underscores.  A proposition can be represented by any identifier that
is not a reserved word.  Connectives and primitives can be represented
in all of the following ways:
\begin{center}
  \begin{tabular}{llll}
    \verb~NOT~     & \verb~!~   & \verb~¬~  &          \\
    \verb~AND~     & \verb~&~   & \verb~&&~ & \verb~∧~ \\
    \verb~OR~      & \verb~|~   & \verb~||~ & \verb~∨~ \\
    \verb~IMPLIES~ & \verb~->~  & \verb~→~  &          \\
    \verb~IFF~     & \verb~<->~ & \verb~↔~  &          \\
    \verb~false~   & \verb~⊥~   &           &          \\
    \verb~true~    & \verb~⊤~   &           &
  \end{tabular}
\end{center}
Operator precedence, from highest to lowest, is: \code{NOT},
\code{AND}, \code{OR}, \code{IMPLIES}, \code{IFF}.  All binary
operators are right-associative.  Of course, parentheses can be used
to group subformulas.

\subsection{Conversion commands}

The commands \code{nnf}, \code{cnf}, \code{dnf}, and \code{tseytin} are
similar in that each converts a given formula to a certain form.  The
usage of \code{cnf} is typical:
\begin{verbatim}
Usage: lap cnf <options> [<filename>]
Description:
  Converts a propositional formula to an equivalent formula in 
  conjunctive normal form.  By default, the formula is read from a
  file, specified by <filename>.  However, using options below, 
  this can be changed to read from stdin or to specify the 
  formula on the command line.  Output is sent to stdout.
Options:
  -in     : read formula from stdin
  -f <string>
          : read the formula from <string> instead of a file
  -v      : verbose output
  -plain  : restrict output to plain text
For formula syntax, type "lap help formulas".
\end{verbatim}
For example:
\begin{verbatim}
> lap cnf -f '!(!p->!q)|r'
(¬p∨r)∧(q∨r) 
\end{verbatim}

\subsection{The \code{dpll} command}

As explained above, \code{dpll} consumes a CNF formula and determines
its satisfiability using the DPLL algorithm.  Option \code{-v} reveals
the algorithmic steps, and \code{-model} displays a satisfying model
at the end if one exists (a model is identified with the set of
propositions it assigns \emph{true}):
\begin{verbatim}
> lap dpll -v -model -f '(p|q)&!p'
CNF formula: (p∨q)∧¬p
CNF Structure: {[¬p], [p,q]}
Push {[¬p], [p,q]}.  Model = {}
[UNIT] Setting p to false.
Push {[q]}.  Model = {}
[UNIT] Setting q to true.
Push {}.  Model = {q}
Satisfying model found!
Pop.
Pop.
Pop.
{q}
\end{verbatim}

\subsection{The  \code{sat} command}

Command \code{sat} determines satisfiability of any propositional
formula, with option \code{-alg} used to specify an algorithm.
Algorithm \code{brute} uses ``brute force,'' iterating over all
models over the propositions occurring in the formula:
\begin{verbatim}
> lap sat -v -model -alg brute -f '(p|q)&!p'
Formula: (p∨q)∧¬p
Evaluating formula at model {}.  Result: false
Evaluating formula at model {p}.  Result: false
Evaluating formula at model {q}.  Result: true
{q}
\end{verbatim}
Algorithm \code{dpll} combines Tseytin's transformation with DPLL, as
described above.  The commands to determine if a formula is
\code{valid} or if two formulas are \code{equiv}alent behave in the
obvious way.

\section{Natural Deduction for Propositional and First Order Logic}

In this section we describe the LAP capability to check natural
deduction derivations for both propositional and first order logic.

\subsection{FOL Syntax}

FOL formula syntax extends the PL formula syntax defined in Section
\ref{sec:formulasyntax}.  A formula may use variables, constants,
function symbols (of arity at least $1$), predicate symbols (of arity
at least $0$), and quantifiers.

LAP does not enforce any convention regarding identifiers used for
predicate or function symbols, variables, or constants.  Any
identifier that is not a reserved word can be used to represent any of
those syntactic elements.  To distinguish variables from constants,
the user must declare the constants at the top of a derivation.
Everything else about the first order vocabulary can be deduced from
context: function and predicate symbols are always followed by
parentheses, the arity is determined by the number of arguments,
predicate applications only occur in a context where a formula is
expected, and function applications in a context where a term is
expected.

The universal and existential quantifiers can be represented as
follows:
\begin{center}
  \begin{tabular}{llll}
    \verb~forall~ & \verb~∀~\\
    \verb~exists~ & \verb~∃~\\
  \end{tabular}
\end{center}
An example of a quantified formula is
\begin{verbatim}
  forall x . P(x) & Q(x)
\end{verbatim}
Note the dot (\code{.}) after the variable (\code{x}).  The
quantifiers have the lowest operator precedence, so the expression
above denotes the same formula as
\begin{verbatim}
  forall x . (P(x) & Q(x))
\end{verbatim}

\subsection{Derivations}

LAP's representation of a PL derivation is based on the following:

\begin{definition}
  \label{def:der}
  An $\natpl$ \emph{derivation} is a triple
  $\mathcal{D} = (s, R, (\mathcal{D}_1, \dots, \mathcal{D}_n))$, written
  \[
    (R) \frac{\mathcal{D}_1 \ \cdots \ \mathcal{D}_n}{s},
  \]
  where $R \in \textup{RULES}_{\textup{PL}}$, $s$ is a sequent (the
  {\emph{conclusion}} of $\mathcal{D}$), $n$ is the number of premises
  of $R$, $\mathcal{D}_i$ is a derivation with conclusion $s_i$
  $(1 \leq i \leq n)$, and
  \[
    (R) \frac{s_1 \ \cdots \ s_n}{s}
  \]
  is an instance of $R$. The $\mathcal{D}_i $ are the \emph{immediate
    subderivations} of $\mathcal{D}$.
\end{definition}

\begin{figure}
  \newcommand{\gap}{\hspace{3ex}}
  \newcommand{\vgap}{4ex}
  \[
    \begin{array}{@{}r@{\;}lr@{\;}l@{}}
      (\Ax)
      & \ds\frac{\phantom{\tsty{A}}}{\mathsf{\tsty{\Gamma}},
        \tsty{A}\vdash \tsty{A}}
      & (\RAA)
      & \ds\frac{\tsty{\Gamma},\neg \tsty{A}\vdash \bot}
        {\tsty{\Gamma}\vdash \tsty{A}}\\[\vgap]
      (\Iimplies)
      & \ds\frac{\tsty{\Gamma}, \tsty{A}\vdash \tsty{B}}
        {\tsty{\Gamma} \vdash \tsty{A}\ra \tsty{B}}
      & (\Inot)
      & \ds\frac{\tsty{\Gamma},\tsty{A}\vdash\bot}
        {\tsty{\Gamma}\vdash\neg \tsty{A}}\\[\vgap]
      (\Iand)
      & \ds\frac{\tsty{\Gamma}\vdash \tsty{A} \gap
        \tsty{\Gamma}\vdash \tsty{B}}
        {\tsty{\Gamma}\vdash \tsty{A}\wedge \tsty{B}}
      & (\Ior{i})
      & \ds\frac{\tsty{\Gamma}\vdash \tsty{A}_i}
        {\tsty{\Gamma} \vdash \tsty{A}_1\vee \tsty{A}_2} \:\:
        (i\in\{1,2\})\\[\vgap]
      (\Eimplies)
      & \ds\frac{\tsty{\Gamma}\vdash \tsty{A} \gap
        \tsty{\Gamma}\vdash \tsty{A}\ra \tsty{B}}
        {\tsty{\Gamma}\vdash \tsty{B}}
      & (\Enot)
      & \ds\frac{\tsty{\Gamma}\vdash \tsty{A} \gap
        \tsty{\Gamma}\vdash\neg \tsty{A}}{\tsty{\Gamma}\vdash
        \tsty{B}}\\[\vgap]
      (\Eand{i})
      & \ds\frac{\tsty{\Gamma}\vdash \tsty{A}_1\wedge \tsty{A}_2}
        {\tsty{\Gamma}\vdash \tsty{A}_i}\:\:
        (i\in\{1,2\})
      & (\Eor)
      & \ds\frac{\tsty{\Gamma}\vdash \tsty{A}\vee \tsty{B}\gap
        \tsty{\Gamma},\tsty{A}\vdash \tsty{C}\gap
        \tsty{\Gamma},\tsty{B}\vdash \tsty{C}}{\tsty{\Gamma} \vdash \tsty{C}}   
    \end{array}
  \]
  \caption{Inference Rules for $\natpl$}
  \label{fig:pl:nat}
\end{figure}

The inference rules ($\textup{RULES}_{\textup{PL}}$) are given in
Figure \ref{fig:pl:nat}.  An $\natfol$ derivation is defined
similarly, using instead $\textup{RULES}_{\textup{FOL}}$, which
contains the four additional rules shown in Figure
\ref{fig:natfol:rules}.

\begin{figure}
  \newcommand{\gap}{\hspace{3ex}}
  \newcommand{\vgap}{4ex}
  \[
    \begin{array}{@{}r@{\;}l@{\hspace{10ex}}l@{}}
      (\Iforall)
      & \ds\frac{\tsty{\Gamma}\vdash\tsty{\phi}[\tsty{y}/\tsty{x}]}{%
        \tsty{\Gamma}\vdash\forall\tsty{x}.\tsty{\phi}}
      & \parbox{.4\textwidth}{$\tsty{y}$ is free for $\tsty{x}$ in $\phi$ and\\
      $\tsty{y}$ does not occur free in $\tsty{\Gamma}$ or
      $\tsty{\phi}$}\\[\vgap]
      (\Eforall)
      & \ds\frac{\tsty{\Gamma}\vdash\forall\tsty{x}.\tsty{\phi}}{%
        \tsty{\Gamma}\vdash\tsty{\phi}[\tsty{t}/\tsty{x}]}
      & \text{$\tsty{t}$ is free for $\tsty{x}$ in $\tsty{\phi}$}
      \\[\vgap]
      (\Iexists)
      & \ds\frac{\tsty{\Gamma}\vdash\tsty{\phi}[\tsty{t}/\tsty{x}]}{%
        \tsty{\Gamma}\vdash\exists\tsty{x}.\tsty{\phi}}
      & \text{$\tsty{t}$ is free for $\tsty{x}$ in $\tsty{\phi}$}
      \\[\vgap]
      (\Eexists)
      & \ds\frac{\tsty{\Gamma}\vdash\exists\tsty{x}.\tsty{\phi}
        \gap
        \tsty{\Gamma},\tsty{\phi}[\tsty{y}/\tsty{x}]\vdash\tsty{\theta}}{%
        \tsty{\Gamma}\vdash\tsty{\theta}}
      & \parbox{.4\textwidth}{$\tsty{y}$ is free for $\tsty{x}$ in $\phi$ and\\
      $\tsty{y}$ does not occur free in $\tsty{\Gamma}$,
      $\tsty{\phi}$, or $\tsty{\theta}$}
    \end{array}
  \]
  \caption{Additional rules for $\natfol$.}
  \label{fig:natfol:rules}
\end{figure}

\subsection{LAP Derivation Syntax}

To enable automated verification, we implement a concrete syntax that
maps these formal constructs to a machine-readable format.

First, the sequent symbol is denoted \code{|-} or \code{⊢}.

The inference rules can be specified using any of the following:
\begin{center}
  \begin{tabular}{lll}
    \verb~Ax~          &                 &                 \\
    \verb~IAND~        & \verb~I&~       & \verb~I∧~       \\
    \verb~EAND1~       & \verb~E&1~      & \verb~E∧1~      \\
    \verb~EAND2~       & \verb~E&2~      & \verb~E∧2~      \\
    \verb~IOR1~        & \verb~I|~       & \verb~I∨1~      \\
    \verb~IOR2~        & \verb~I|2~      & \verb~I∨2~      \\
    \verb~EOR~         & \verb~E|~       & \verb~E∨~       \\
    \verb~IIMPLIES~    & \verb~I->~      & \verb~I→~       \\
    \verb~EIMPLIES~    & \verb~E->~      & \verb~E→~       \\
    \verb~INOT~        & \verb~I!~       & \verb~I¬~       \\
    \verb~ENOT~        & \verb~E!~       & \verb~E¬~       \\
    \verb~RAA~         &                 &                 \\
    \verb~Iforall~     & \verb~IA~       & \verb~I∀~       \\
    \verb~Eforall~     & \verb~EA~       & \verb~E∀~       \\
    \verb~Iexists~     & \verb~IE~       & \verb~I∃~       \\
    \verb~Eexists~     & \verb~EE~       & \verb~E∃~       \\
  \end{tabular}
\end{center}

A derivation is expressed in a linear format, as a sequence of steps.
Each step begins with a number, followed by a dot (\code{.}).  This is
followed by a sequent: a comma-separated list of formulas, followed by
the sequent symbol, followed by a formula.  This is followed by the
name of a rule in parentheses, e.g., \code{(RAA)}.  If the rule has
premises, this is followed by a comma-separated list of numbers, the
line numbers of the premises.  Finally, the step is terminated by a
dot.  As with formulas, white space is ignored.

In summary, a step has the form:
\begin{verbatim}
  <number>. <sequent> (<rule>) <premises>.
\end{verbatim}
An example step is:
\begin{verbatim}
  3. p, p->q |- q (E->)1,2.
\end{verbatim}

As mentioned above, an FOL derivation that uses constants requires
that those constants first be declared.  These are specified using the
\code{const} keyword, followed by a comma-separated list of
identifiers and a semicolon, e.g.:
\begin{verbatim}
  const a, b, c;
\end{verbatim}

Here is a complete example of an FOL derivation that is accepted by
LAP:
\begin{verbatim}
  const c;
  1. forall x. P(x) |- forall x. P(x) (Ax).
  2. forall x. P(x) |- P(c) (Eforall)1.
  3. forall x. P(x) |- exists y. P(y) (Iexists)2.
\end{verbatim}
Using Unicode characters, the same derivation could be written
\begin{verbatim}
  const c;
  1. ∀x.P(x) ⊢ ∀x.P(x) (Ax).
  2. ∀x.P(x) ⊢ P(c)    (E∀)1.
  3. ∀x.P(x) ⊢ ∃y.P(y) (I∃)2.
\end{verbatim}

Finally, the \code{\#} character is used to start a comment that
extends to the end of the line.

\subsection{Output Formats}

While the input format is currently restricted to the linear
form described above, LAP has the ability to output the derivation
in several different formats, or \emph{views}:
\begin{itemize}
\item \emph{tuple}: the raw mathematical representation, using nested
  parentheses, based on Definition \ref{def:der};
\item \emph{linear}: the input format described above;
\item \emph{fitch}: a Fitch diagram, which uses nested scopes to specify the
  context;
\item \emph{tree}: a traditional mathematical layout with the root at the
  bottom, the second representation shown in Definition \ref{def:der};
  and
\item \emph{hierarchy}: an indented directory-style tree.
\end{itemize}

While they can look quite different, all of these views are just different
ways of displaying the same underlying mathematical object.  To really
emphasize this point, LAP has the ability to number the conclusions
(or subderivations) in every view, using the same numbering scheme in
all cases.  This allows a user to see exactly how a particular node in
the tree view, for example, corresponds to a line in the Fitch
diagram.
 
\subsection{Checking a Derivation}

\begin{figure}
\begin{lap}
> lap check -lang fol -view all -number quantifierdual.lap
true

(\textcolor{blue}{⑦} ∀x.¬P(x) ⊢ ¬∃x.P(x), \textcolor{red}{I¬}, (
  (\textcolor{blue}{⑥} ∀x.¬P(x),∃x.P(x) ⊢ ⊥, \textcolor{red}{E∃}, (
    (\textcolor{blue}{①} ∀x.¬P(x),∃x.P(x) ⊢ ∃x.P(x), \textcolor{red}{Ax}, ()), 
    (\textcolor{blue}{⑤} P(y),∀x.¬P(x),∃x.P(x) ⊢ ⊥, \textcolor{red}{E¬}, (
      (\textcolor{blue}{②} P(y),∀x.¬P(x),∃x.P(x) ⊢ P(y), \textcolor{red}{Ax}, ()), 
      (\textcolor{blue}{④} P(y),∀x.¬P(x),∃x.P(x) ⊢ ¬P(y), \textcolor{red}{E∀}, (
        (\textcolor{blue}{③} P(y),∀x.¬P(x),∃x.P(x) ⊢ ∀x.¬P(x), \textcolor{red}{Ax}, ())))))))))

\textcolor{blue}{1.}      ∀x.¬P(x),∃x.P(x) ⊢ ∃x.P(x)  \textcolor{red}{(Ax)}\textcolor{blue}{.}   
\textcolor{blue}{2.} P(y),∀x.¬P(x),∃x.P(x) ⊢ P(y)     \textcolor{red}{(Ax)}\textcolor{blue}{.}   
\textcolor{blue}{3.} P(y),∀x.¬P(x),∃x.P(x) ⊢ ∀x.¬P(x) \textcolor{red}{(Ax)}\textcolor{blue}{.}   
\textcolor{blue}{4.} P(y),∀x.¬P(x),∃x.P(x) ⊢ ¬P(y)    \textcolor{red}{(E∀)}\textcolor{blue}{3.}  
\textcolor{blue}{5.} P(y),∀x.¬P(x),∃x.P(x) ⊢ ⊥        \textcolor{red}{(E¬)}\textcolor{blue}{2,4.}
\textcolor{blue}{6.}      ∀x.¬P(x),∃x.P(x) ⊢ ⊥        \textcolor{red}{(E∃)}\textcolor{blue}{1,5.}
\textcolor{blue}{7.}              ∀x.¬P(x) ⊢ ¬∃x.P(x) \textcolor{red}{(I¬)}\textcolor{blue}{6.}  

\textcolor{blue}{7. }\textcolor{blue}{}∀x.¬P(x) ⊢ ¬∃x.P(x)\textcolor{red}{  (I¬)}
\textcolor{blue}{6. }\textcolor{blue}{│ }∀x.¬P(x),∃x.P(x) ⊢ ⊥\textcolor{red}{  (E∃)}
\textcolor{blue}{1. }\textcolor{blue}{│ │ }∀x.¬P(x),∃x.P(x) ⊢ ∃x.P(x)\textcolor{red}{  (Ax)}
\textcolor{blue}{5. }\textcolor{blue}{│ │ }P(y),∀x.¬P(x),∃x.P(x) ⊢ ⊥\textcolor{red}{  (E¬)}
\textcolor{blue}{2. }\textcolor{blue}{│ │ │ }P(y),∀x.¬P(x),∃x.P(x) ⊢ P(y)\textcolor{red}{  (Ax)}
\textcolor{blue}{4. }\textcolor{blue}{│ │ │ }P(y),∀x.¬P(x),∃x.P(x) ⊢ ¬P(y)\textcolor{red}{  (E∀)}
\textcolor{blue}{3. }\textcolor{blue}{│ │ │ │ }P(y),∀x.¬P(x),∃x.P(x) ⊢ ∀x.¬P(x)\textcolor{red}{  (Ax)}

   \textcolor{blue}{│ }∀x.¬P(x)            
   \textcolor{blue}{├───}                  
   \textcolor{blue}{│ │ }∃x.P(x)           
   \textcolor{blue}{│ ├───}                
\textcolor{blue}{1.} \textcolor{blue}{│ │ }∃x.P(x)    \textcolor{red}{(Ax)}\textcolor{blue}{}  
   \textcolor{blue}{│ │ │ }P(y)            
   \textcolor{blue}{│ │ ├───}              
\textcolor{blue}{2.} \textcolor{blue}{│ │ │ }P(y)     \textcolor{red}{(Ax)}\textcolor{blue}{}  
\textcolor{blue}{3.} \textcolor{blue}{│ │ │ }∀x.¬P(x) \textcolor{red}{(Ax)}\textcolor{blue}{}  
\textcolor{blue}{4.} \textcolor{blue}{│ │ │ }¬P(y)    \textcolor{red}{(E∀)}\textcolor{blue}{3}  
\textcolor{blue}{5.} \textcolor{blue}{│ │ │ }⊥        \textcolor{red}{(E¬)}\textcolor{blue}{2,4}
\textcolor{blue}{6.} \textcolor{blue}{│ │ }⊥          \textcolor{red}{(E∃)}\textcolor{blue}{1,5}
\textcolor{blue}{7.} \textcolor{blue}{│ }¬∃x.P(x)     \textcolor{red}{(I¬)}\textcolor{blue}{6}  
\end{lap}
\caption{Derivation of
  $\forall x. \neg P(x) \vdash \neg \exists x. P(x)$.  With
  \code{-view all}, after checking the derivation and reporting it is
  correct, LAP also prints the derivation in 5 views: tuple, linear,
  hierarchy, fitch, and tree. The tree view is shown in Figure
  \ref{fig:SC-quantifierdual-tree}.}
\label{fig:SC-quantifierdual}
\end{figure}

\begin{figure}
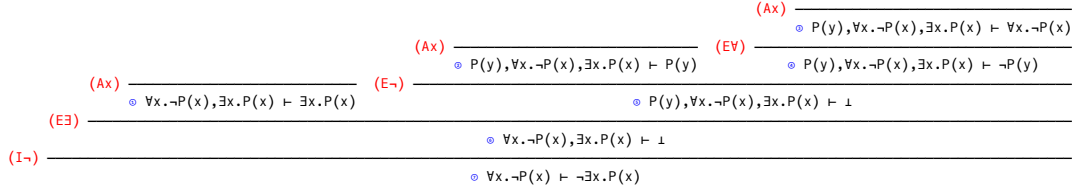

  \centering
\begin{tinylap}
                                                                                            \textcolor{red}{(Ax)} ──────────────────────────────────
                                                                                                 \textcolor{blue}{③} P(y),∀x.¬P(x),∃x.P(x) ⊢ ∀x.¬P(x)
                                                  \textcolor{red}{(Ax)} ──────────────────────────────  \textcolor{red}{(E∀)} ───────────────────────────────────────
                                                       \textcolor{blue}{②} P(y),∀x.¬P(x),∃x.P(x) ⊢ P(y)           \textcolor{blue}{④} P(y),∀x.¬P(x),∃x.P(x) ⊢ ¬P(y)
          \textcolor{red}{(Ax)} ────────────────────────────  \textcolor{red}{(E¬)} ─────────────────────────────────────────────────────────────────────────────────
               \textcolor{blue}{①} ∀x.¬P(x),∃x.P(x) ⊢ ∃x.P(x)                                  \textcolor{blue}{⑤} P(y),∀x.¬P(x),∃x.P(x) ⊢ ⊥
     \textcolor{red}{(E∃)} ─────────────────────────────────────────────────────────────────────────────────────────────────────────────────────────
                                                           \textcolor{blue}{⑥} ∀x.¬P(x),∃x.P(x) ⊢ ⊥
\textcolor{red}{(I¬)} ──────────────────────────────────────────────────────────────────────────────────────────────────────────────────────────────
                                                         \textcolor{blue}{⑦} ∀x.¬P(x) ⊢ ¬∃x.P(x)
\end{tinylap}
  \caption{Derivation of $\forall x. \neg P(x) \vdash \neg \exists x. P(x)$, tree view.}
  \label{fig:SC-quantifierdual-tree}
\end{figure}

As the derivation is parsed, LAP constructs its internal
representation, a derivation structure.  Each time a new derivation
instance is about to be created, a check occurs to ensure the rule of
the derivation has been correctly applied, including the side
conditions, if any.  If the final (root) derivation is correct, the
program will print \code{true}, and optionally output one or more
views of this derivation.

An example terminal interaction to check a valid derivation is shown
in Figures \ref{fig:SC-quantifierdual} and
\ref{fig:SC-quantifierdual-tree}.  The \code{-number} option causes
LAP to include numbers in the tuple, tree, and hierarchy views
(numbers are always included in the linear and Fitch views).  With
this option, it is easy to see how each view represents the same
structure, which in this case is essentially a tree with 7 nodes.
Note that these figures are \emph{not} screenshots, but text that can
be easily copied and pasted.

\subsection{Error Explanations}

If any check fails, LAP will output \code{false} and optionally 
an error explanation.  As an example, consider the following
erroneous derivation:
\begin{verbatim}
1.   p|q |- p|q (Ax).     
2. p|q,p |- p   (Ax).     
3. p|q,p |- q|p (IOR1)2.     # erroneous application of IOR1!
4. p|q,q |- q   (Ax).     
5. p|q,q |- q|p (IOR1)4.   
6.   p|q |- q|p (EOR)1,3,5.
\end{verbatim}

\begin{figure}[t]
\begin{lap}
> lap check -lang pl -v badOr1.lap      
false
Violation of rule I∨1 at step 3:
  Premise 1  : p∨q,p ⊢ p
  Conclusion : p∨q,p ⊢ q∨p
The premise's succedent, p, should be the OR formula's left argument, q,
but is not.

Rule I∨1 ("introduce or 1"):
      Γ ⊢ A
     ───────
     Γ ⊢ A∨B
Rule I∨1 says that if you know A, then you can conclude A∨B. The premise
and the conclusion use the same context Γ. This rule has one premise.
\end{lap}
\caption{The output of an example with $\Ior{1}$ rule violation.}
\label{fig:badOr1.lap}
\end{figure}

\begin{figure}
\begin{lap}
> lap check -lang fol -v badEexists1.lap
false
Violation of rule E∃ at step 3:
  Premise 1  : ∃x.P(x) ⊢ ∃x.P(x)
  Premise 2  : P(y),∃x.P(x) ⊢ P(y)
  Conclusion : ∃x.P(x) ⊢ P(y)
This step violates Side condition 1 as y does occur free in the
conclusion's succedent θ (P(y))

Rule E∃ ("eliminate exists") :
    Γ ⊢ ∃x φ    Γ, φ[y/x] ⊢ θ
    ─────────────────────────
              Γ ⊢ θ
Side condition 1: y must not occur free in Γ, φ, or θ;
Side condition 2: y must be free for x in φ.
\end{lap}
\caption{The output of an example with $\Eexists$ side condition violation.}
\label{fig:SC-bad_eexists_variable_leak}
\end{figure}

Figure \ref{fig:badOr1.lap} shows the result of checking this
derivation with LAP.  The explanation states the exact step at which
the error occurs.  It shows that the user is trying to apply rule
$\Ior{1}$ and shows the user's premise and conclusion.  It then
explains exactly why this is not a valid instance of the rule: the
\code{p} should occur as the left argument of the \emph{or} operator
in the concluding formula \code{q∨p}, but instead the left argument is
\code{q}.  It then provides a specification and informal description
of the rule, for reference.  The user would then proceed to correct
their derivation and re-run LAP.

For another example of a violation, consider the candidate derivation
\begin{verbatim}
1. ∃x. P(x) ⊢ ∃x. P(x) (Ax).
2. ∃x. P(x), P(y) ⊢ P(y) (Ax).
3. ∃x. P(x) ⊢ P(y) (E∃)1,2.     # erroneous application of E∃!
\end{verbatim}
The output from LAP is shown in Figure
\ref{fig:SC-bad_eexists_variable_leak}.  In this case, one of the side
conditions of rule $\Eexists$ is violated.

\section{System Design and Future Work}

\begin{figure}
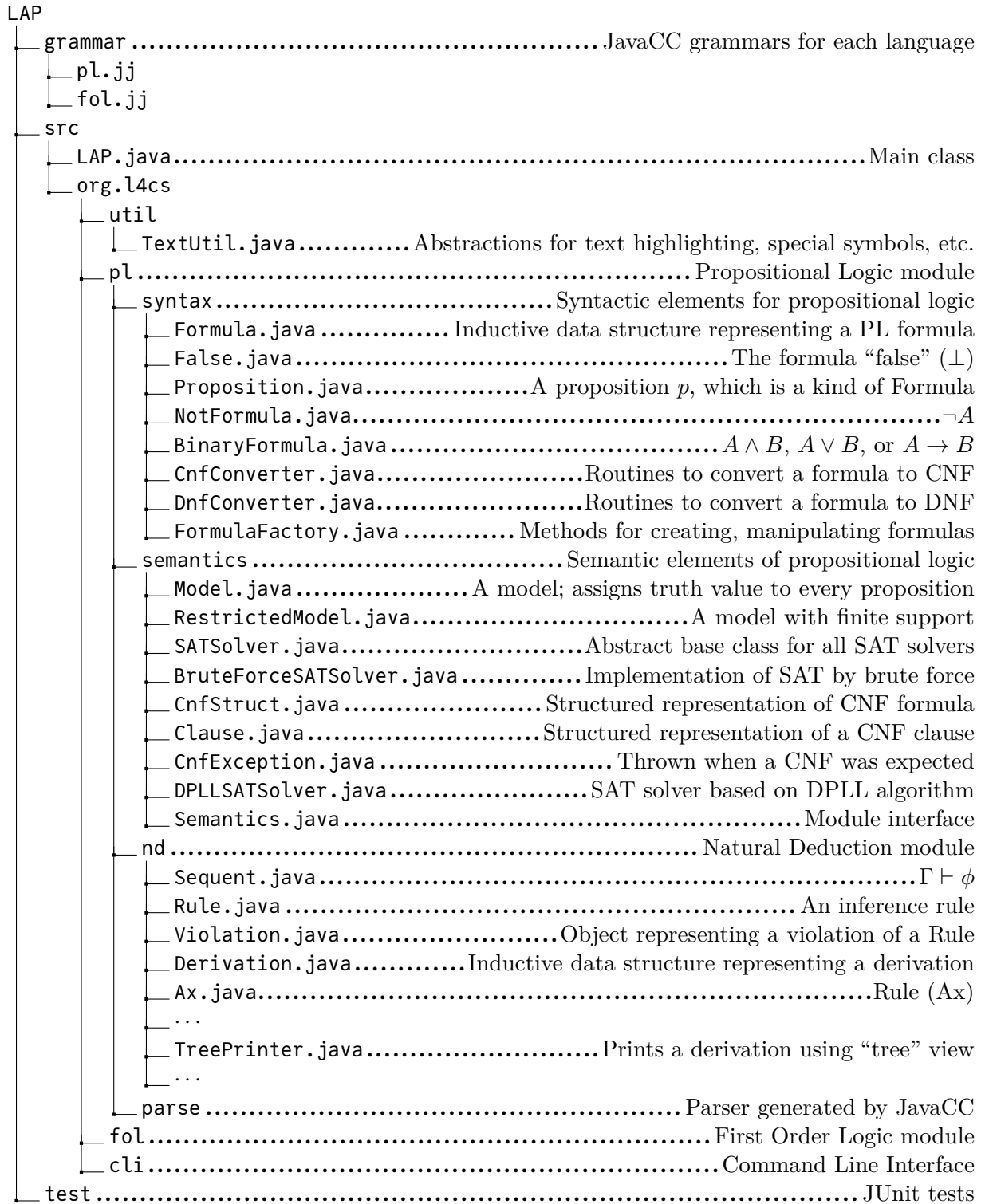

  \dirtree{%
    .1 LAP.
    .2 grammar\DTcomment{JavaCC grammars for each language}.
    .3 pl.jj.
    .3 fol.jj.
    .2 src.
    .3 LAP.java\DTcomment{Main class}.
    .3 org.l4cs.
    .4 util.
    .5 TextUtil.java\DTcomment{Abstractions for text highlighting,
      special symbols, etc.}.
    .4 pl\DTcomment{Propositional Logic module}.
    .5 syntax\DTcomment{Syntactic elements for propositional logic}.
    .6 Formula.java\DTcomment{Inductive data structure representing a PL formula}.
    .6 False.java\DTcomment{The formula ``false'' ($\bot$)}.
    .6 Proposition.java\DTcomment{A proposition $p$, which is a kind
      of Formula}.
    .6 NotFormula.java\DTcomment{$\neg A$}.
    .6 BinaryFormula.java\DTcomment{$A\wedge B$, $A\vee B$, or $A\ra B$}.
    .6 CnfConverter.java\DTcomment{Routines to convert a formula to CNF}.
    .6 DnfConverter.java\DTcomment{Routines to convert a formula to DNF}.
    .6 FormulaFactory.java\DTcomment{Methods for creating, 
      manipulating formulas}. 
    .5 semantics\DTcomment{Semantic elements of propositional logic}.
    .6 Model.java\DTcomment{A model; assigns truth value to every proposition}.
    .6 RestrictedModel.java\DTcomment{A model with finite support}.
    .6 SATSolver.java\DTcomment{Abstract base class for all SAT solvers}. 
    .6 BruteForceSATSolver.java\DTcomment{Implementation of SAT 
      by brute force}. 
    .6 CnfStruct.java\DTcomment{Structured representation of CNF formula}.
    .6 Clause.java\DTcomment{Structured representation of a CNF clause}.
    .6 CnfException.java\DTcomment{Thrown when a CNF was expected}.
    .6 DPLLSATSolver.java\DTcomment{SAT solver based on DPLL algorithm}.
    .6 Semantics.java\DTcomment{Module interface}.
    .5 nd\DTcomment{Natural Deduction module}.
    .6 Sequent.java\DTcomment{$\Gamma\infers\phi$}.
    .6 Rule.java\DTcomment{An inference rule}.
    .6 Violation.java\DTcomment{Object representing a violation of a Rule}.
    .6 Derivation.java\DTcomment{Inductive data structure
      representing a derivation}.
    .6 Ax.java\DTcomment{Rule (Ax)}.
    .6 $\cdots$.
    .6 TreePrinter.java\DTcomment{Prints a derivation using ``tree''
      view}.
    .6 $\cdots$.
    .5 parse\DTcomment{Parser generated by JavaCC}.
    .4 fol\DTcomment{First Order Logic module}.
    .4 cli\DTcomment{Command Line Interface}.
    .2 test\DTcomment{JUnit tests}.
  }
  \caption{Directory structure of LAP source code.}
  \label{fig:design}
\end{figure}

LAP is composed of modules with clearly-defined interfaces.  The
directory structure of the LAP source code, which also reflects the
submodule (or \emph{is-component-of}) relation, is shown, in part, in
Figure \ref{fig:design}.  The \emph{uses} relation has also been
carefully constrained.  For example, the submodules of \code{pl}
satisfy: \code{syntax} does not use any other module, \code{semantics}
uses only \code{syntax}, \code{nd} uses only \code{syntax}, and
\code{parse} uses only \code{syntax} and \code{nd}.  In particular,
there are no cycles in that \emph{uses} relation.  The \code{pl} and
\code{fol} modules are completely independent, which results in some
duplication, but also makes the \code{pl} module much easier to
understand, especially for a beginning student.

Together with thorough documentation, we think these architectural
decisions make the code easy to understand, maintain, and extend.  We
hope this will encourage others to use and contribute to the project.

For those interested in contributing, here are some ideas for future
work:
\begin{itemize}
\item For both propositional and first order logic: add some derived
  rules, and show how a derivation using them is transformed to one
  that does not.
\item Support first order logic with equality.
\item Explore some first order theories.
\item Add transformation to Herbrand and Skolem normal forms.
\item Support Herbrand models, and implement a semi-decision procedure
  for validity of first order formulas.
\item Add more languages, such as Hoare logic (for a simple
  programming language) and temporal logics.
\end{itemize}

\subsection*{Acknowledgments}

This material is based upon work supported by the U.S.\ National
Science Foundation under Award Number CCF-2446130, and by the U.S.\
Department of Energy, Office of Science, Advanced Scientific Computing
Research (ASCR) Program, under Award Number DE-SC0025953.

This report was prepared as an account of work sponsored by an agency
of the United States Government. Neither the United States Government
nor any agency thereof, nor any of their employees, makes any
warranty, express or implied, or assumes any legal liability or 
responsibility for the accuracy, completeness, or usefulness of any
information, apparatus, product, or process disclosed, or represents
that its use would not infringe privately owned rights. Reference
herein to any specific commercial product, process, or service by
trade name, trademark, manufacturer, or otherwise does not necessarily
constitute or imply its endorsement, recommendation, or favoring by
the United States Government or any agency thereof. The views and
opinions of authors expressed herein do not necessarily state or
reflect those of the United States Government or any agency thereof.

\bibliography{lap}

\end{document}